# Effect of secondary LuNiSn phase on thermoelectric properties of half-Heusler alloy LuNiSb

Karol Synoradzki1, Kamil Ciesielski, Leszek Kępiński, Dariusz Kaczorowski

*Institute of Low Temperature and Structure Research, Polish Academy of Sciences, Okólna 2, 50-422 Wrocław, Poland*

**Abstract**

We report on the high-temperature (350 K < $T$ < 1000 K) thermoelectric properties of the ternary compounds LuNiSb (cubic) and LuNiSn (orthorhombic) and a composite material $(LuNiSb)_{0.5}(LuNiSn)_{0.5}$. The electrical conductivity in LuNiSn is metallic, while it is semiconducting-like in LuNiSb. The Seebeck coefficient reaches -5.5 µV/K at 700 K for the former compound and 66 µV/K at 607 K for the latter one. The composite sample $(LuNiSb)_{0.5}(LuNiSn)_{0.5}$ constructed from orthorhombic matrix with cubic inclusions combines the electrical conductivity of LuNiSn with the thermoelectric properties of LuNiSb. Nonetheless, no enhancement of the thermoelectric performance occurs for this material.

## 1. Introduction

Half-Heusler (HH) alloys, crystallizing in the cubic MgAgAs-type structure (space group *F-43m*, No. 216), are being widely investigated due to their promising potential for thermoelectric (TE) power generation applications. In particular, transition metal *n*-type stannides *M*NiSn and *p*-type antimonides *M*CoSb (*M* = Zr, Hf, Ti) have attracted tremendous interest in recent years [1–5]. Much less studied remain the corresponding rare-earth (RE) based HH alloys, which have also been attested to show prospective TE characteristics [6–17]. In order to achieve even better TE properties, some modification could be realized. A well-established method for improving TE performance is introduction of an additional phase to a given thermoelectric material. Secondary phase inclusions may act not only as defect centers for phonon scattering but also as low energy carriers filters [18,19]. Moreover, composite samples can combine physical properties of their constituent compounds, which may lead to improvement of the TE effect.

The main goal of this work, which is a part of our broader study on TE materials, was to investigate thermoelectric properties of the HH antimonide LuNiSb [20–22]. The compound is a Pauli paramagnet that exhibits semiconducting-like electrical resistivity and large positive Seebeck coefficient (*S*) of 136 µV/K near room temperature [20]. However, in the literature one finds also diverse data on the thermoelectric power of this material, like positive *S* = 70 µV/K at 380 K [21] and negative *S* = –48 µV/K at 300 K [22]. The scatter may arise from structural disorder present in the LuNiSb samples investigated by different authors, and our aim was to check the reproducibility of electrical transport characteristics on different samples of this alloy. Furthermore, we intended to improve the TE performance of LuNiSb by preparing a composite with LuNiSn. This research was driven by the theoretical prediction [23] of notable enhancement of the TE power factor in composites made of metal grains surrounded by a good thermoelectric material. Similar approach applied for some transition metal based HH phases, with metallic full-Heusler phases addition, has resulted in sizable improvement of their TE properties [24–27]. The metallic constituent selected, i.e., LuNiSn, crystallizes in the orthorhombic TiNiSi-type structure (space group *Pnma*, No. 62) [28,29]. Alike LuNiSb it is a Pauli paramagnet, yet its electrical conductance has a metallic character, and its thermopower is small and negative (*S* = −7.2 µV/K) at room temperature [29]. Our selection of LuNiSn being an ingredient in the LuNiSb-based composite was motivated by the fact of (i) its metallic transport properties, (ii) similar chemical composition, (iii) different crystal structure with distinctly different lattice parameters. It is worth noting, that no data on TE behavior of LuNiSn at elevated temperatures are available in the literature.

* Corresponding author. Tel.: +48 713 954 305.
  *E-mail address:* k.synoradzki@int.pan.wroc.pl

In this paper, we report on the synthesis of the HH alloy LuNiSb, the stannide LuNiSn, and the composite (LuNiSb)$_{0.5}$(LuNiSn)$_{0.5}$ as well as on the thermoelectric properties of these materials at temperatures 350 K $<T<$ 1000 K.

## 2. Experimental details

Polycrystalline alloys of LuNiSb, LuNiSn and (LuNiSb)$_{0.5}$(LuNiSn)$_{0.5}$ were prepared by arc-melting stoichiometric amounts of the constituent elements (Lu 99.9%, Ni 99.999 %, Sn 99.99 %, Sb 99.99 %) on a water-cooled copper hearth. The samples were remelted several times in Ti-gettered argon atmosphere. The final mass losses were smaller than 1%. No further heat treatment was applied. For the sake of reproducibility study, two different samples of LuNiSb (hereafter labeled No. 1 and No. 2) were prepared.

Sample characterization was performed at room temperature by means of powder X-ray diffraction using an Xpert Pro PANalytical diffractometer analysis with Cu K$\alpha$ radiation. Quantitative Rietveld refinement was carried out using the Fullprof software to determine lattice parameters, average crystalline size $d$, and phase distribution. To calculate $d$, we used the microstructural analysis from Fullprof. Instrumental resolution file (IRF) of the diffractometer was obtained employing $CeO_2$ as a calibration sample. Chemical composition and distribution of the constituent elements in the samples prepared were determined using a FEI Nova NanoSEM 230 FE-scanning electron microscope (SEM) equipped with a Genesis XM4 energy-dispersive spectrometer (EDS).

Bar-shaped specimens for electrical transport studies were cut from the polycrystalline ingots using a wire saw. Electrical resistivity and thermoelectric power measurements were performed simultaneously in He atmosphere using a Linseis LSR-3 device. Pure platinum was adapted as a reference material. The estimated experimental uncertainties were 3% and 5%, respectively. The measurements were made in the temperature range 350-1000 K.

## 3. Results and discussion

The XRD patterns for all the investigated materials are presented in Fig.1a together with the calculated Bragg peaks positions corresponding to the TiNiSi- and MgAgAs-type structures. The XRD results confirmed that the pristine LuNiSb and LuNiSn alloys crystallize with the cubic *F-43m* and orthorhombic *Pnma* unit cells, respectively. The diffractogram of the composite sample has primary peaks that retain the characteristic peaks of LuNiSb as well as those of LuNiSn. This finding proves the formation of a composite material (LuNiSb)$_{0.5}$(LuNiSn)$_{0.5}$. Each sample, however, contained small amounts of impurity phases such as $Lu_2O_3$ or NiSn/NiSb (indicated in Fig.1a by asterisks).

The lattice parameters of the alloys investigated are shown at Fig. 1b as a function of the content x of the orthorhombic phase LuNiSn. The index x refers to the TiNiSi-type phase weight fraction determined from XRD by the Rietveld method. The lattice parameters derived for LuNiSn are in good agreement with the literature data [28,29]. For the two prepared alloys of LuNiSb, the lattice parameter $a_{HH}$ is slightly smaller than the values reported in the literature [20,21], and it is somewhat sample-dependent. Remarkably, for the composite sample, the lattice parameters are different than those of the constituent compounds. This effect might be related to different level of strain and site disorder in different samples. The average crystalline sizes $d$ (see Table 1) are similar in the samples of LuNiSb and (LuNiSb)$_{0.5}$(LuNiSn)$_{0.5}$, while $d$ is twice larger in the sample of LuNiSn.

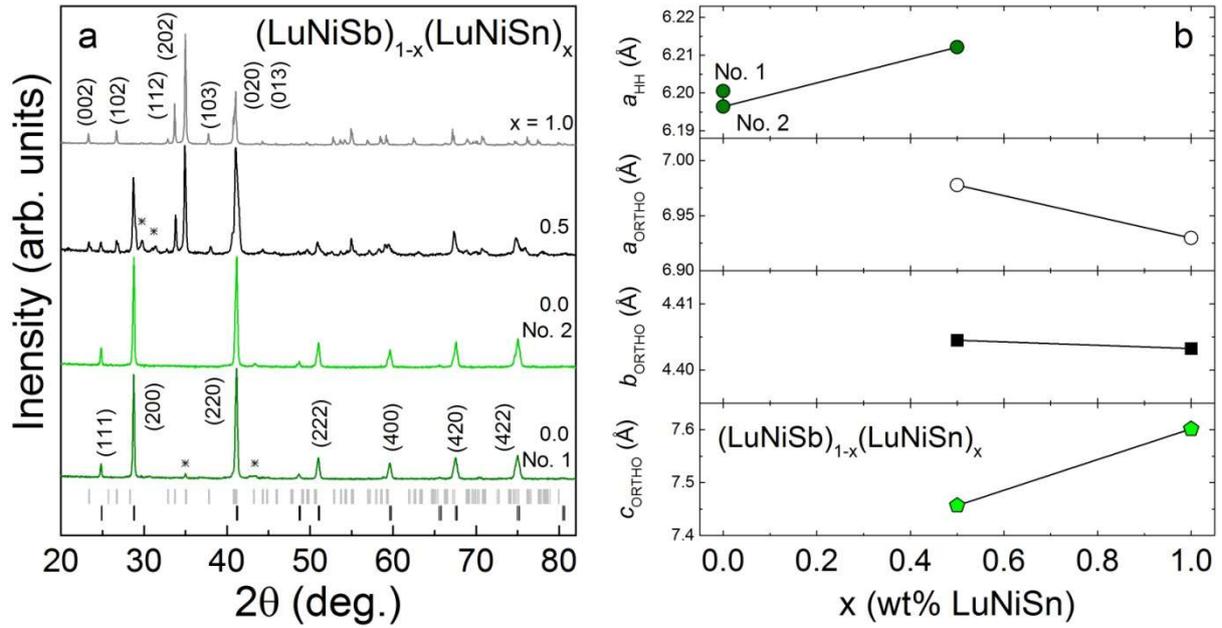

Fig. 1. (a) XRD pattern of the $(LuNiSb)_{1-x}(LuNiSn)_x$ samples. The upper and lower ticks represent Bragg positions corresponding to the orthorhombic and cubic phases, respectively; (b) crystal structure parameters obtained from refinement of the XRD patterns, $a_{HH}$ - lattice parameter of half-Heusler phase, $a_{ORTHO}$, $b_{ORTHO}$, and $c_{ORTHO}$ - lattice parameters of orthorhombic phase.

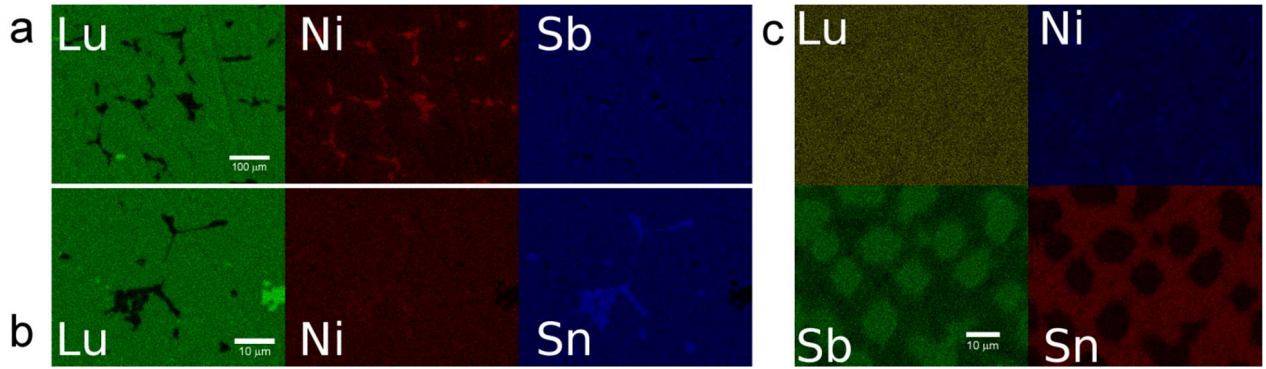

Fig. 2. EDS maps for (a) LuNiSb No. 2, (b) LuNSn and (c) $(LuNiSb)_{0.5}(LuNiSn)_{0.5}$.

The metallographic pictures (not shown here) revealed for each sample investigated minor amount of pores, voids and cracks. The average chemical compositions obtained from a broad surface area studied by EDS are listed in Table 1. For the composite sample, the estimated composition is in good agreement with the nominal one, whereas for the parent compounds an excess of Lu and some Sb/Sn deficiency is observed. Figs. 2a-c present the EDS maps of the specimens investigated. For LuNiSb and LuNiSn, binary phases NiSb/NiSn are seen as impurity phases located at grain boundaries. Remarkably, the composite sample shows clear phase segregation with its matrix made of LuNiSn-like phase and inclusions of LuNiSb. Furthermore, in the orthorhombic matrix phase, some of tin atoms are substituted by antimony atoms. The average size of the LuNiSb inclusions is about 10 μm, which seems larger than a nanoscale phonon mean free path. The latter finding implies that the inclusions may not be very effective for phonon scattering and hence lowering the thermal conductivity.

Table 1. Microstructural parameters for $(LuNiSb)_{1-x}(LuNiSn)_x$ materials, $d$ - average crystalline size.

| Sample | nominal composition Lu:Ni:Sb:Sn | EDS estimated composition Lu:Ni:Sb:Sn | amount of orthorhombic phase | $d$ (Å) |
|---|---|---|---|---|
| LuNiSb No. 1 | 33.3:33.3:33.3:0.0 | 35:34:31:0.0 | 0.0 | 348(49) |
| LuNiSb No. 2 | 33.3:33.3:33.3:0.0 | 43:31:26:0.0 | 0.0 | 347(59) |
| $(LuNiSb)_{0.5}(LuNiSn)_{0.5}$ | 33.3:33.3:16.65:16.65 | 34:34:16:16 | 0.50(1) | 347(71) |
| LuNiSn | 33.3:33.3:0.0:33.3 | 43:31:0.0:26 | 1.0 | 653(98) |

Fig. 3a shows the temperature variation of the electrical resistivity, $\rho(T)$, of LuNiSb, LuNiSn and $(LuNiSb)_{0.5}(LuNiSn)_{0.5}$. The latter two samples show metallic behavior. For LuNiSn, the $\rho$ value at 380 K is of the order of 1 µΩm, in reasonable agreement with the literature data [29]. For the composite sample, the $\rho$ value is twice larger, which can be attributed to crystallographic disorder and secondary phase inclusions. In turn, the electrical resistivity of the two LuNiSb samples investigated varies with temperature in a semiconducting–like manner (no difference between samples No.1 and No.2 was observed).

The temperature dependencies of the Seebeck coefficient, $S(T)$, measured for the four samples examined are presented in Fig.3b. The maximum absolute $S$ values are 66, 54, 18, and -5.5 µV/K recorded at 607, 642, 1000, and 703 K for LuNiSb No.1, LuNiSb No.2, $(LuNiSb)_{0.5}(LuNiSn)_{0.5}$ and LuNiSn, respectively. For the latter alloy, a negative thermopower with an extreme around 700 K is seen. The $S$ value near room temperature equals −3.6 µV/K, i.e. significantly differs from that reported before ($S$ = -7.2 µV/K [29]). In the case of LuNiSb, the Seebeck coefficient is positive and $S(T)$ forms a maximum near 600 or 650 K, depending on the sample. Close to room temperature, the magnitude of $S$ is about 2.5 times smaller than the value given in Ref. [22], yet similar to that quoted in Ref. [21]. The scatter in the observed values likely arises because of some structural disorder or/and nonstoichiometry in the different samples measured. For the composite alloy $(LuNiSb)_{0.5}(LuNiSn)_{0.5}$, the Seebeck coefficient increases linearly with increasing temperature. The thermoelectric power is positive, which suggests that hole-type carriers dominate the electrical transport in this material. The positive sign of $S$ can be related to the LuNiSb inclusions. It seems also possible that the orthorhombic LuNiSn-like matrix changes its overall character of the charge transport from electron- to hole-type upon minor chemical substitution of Sb for Sn encountered for $(LuNiSb)_{0.5}(LuNiSn)_{0.5}$ (see above).

Using the measured $\rho(T)$ and $S(T)$ data, the power factor, PF = $S^2/\rho$, can be calculated, and the results are displayed in Fig.3c. The highest PF value of $0.49 \times 10^{-3}$ Wm$^{-1}$ K$^{-2}$ at 647 K was obtained for LuNiSb No.1, but this value is smaller than PF reported for a few other RE-based HH compounds [12–14,17]. For the sample of LuNiSn, the maximum PF value is as small as $0.02 \times 10^{-3}$ Wm$^{-1}$ K$^{-2}$. In turn, for the composite alloy $(LuNiSb)_{0.5}(LuNiSn)_{0.5}$, PF is a linear function of temperature and reaches $0.08 \times 10^{-3}$ Wm$^{-1}$ K$^{-2}$ at 1000 K.

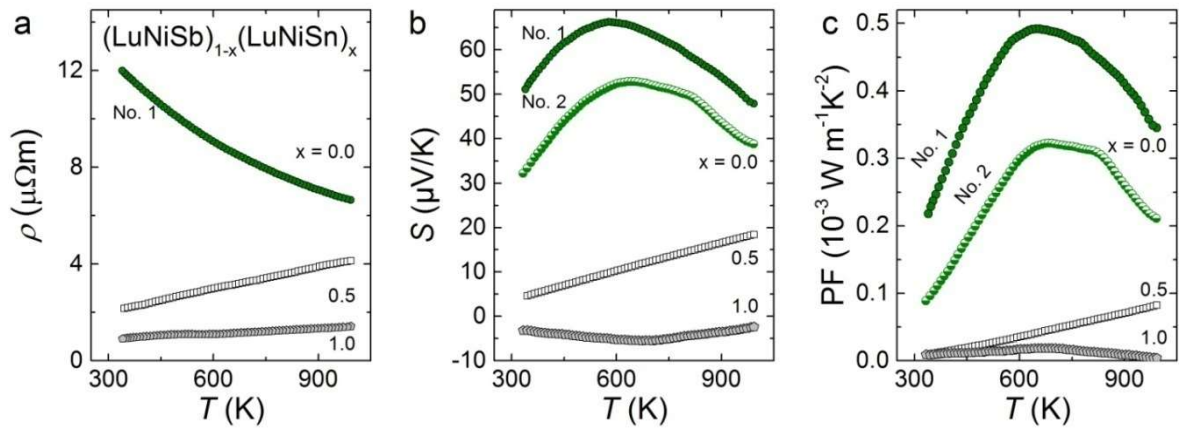

Fig. 3. Temperature dependencies of the electrical resistivity (a), Seebeck coefficient (b), and power factor for $(LuNiSb)_{1-x}(LuNiSn)_x$ samples.

## 4. Summary

We investigated the high-temperature (350 K <$T$< 1000 K) thermoelectric properties of LuNiSb, LuNiSn and the composite material $(LuNiSb)_{0.5}(LuNiSn)_{0.5}$. The half-Heusler alloy LuNiSb was found to exhibit a semiconducting-like resistivity and positive Seebeck coefficient. In turn, the orthorhombic compound LuNiSn was established to show a metallic conductivity with negative thermopower. The composite sample (LuNiSb)$_{0.5}$(LuNiSn)$_{0.5}$, constructed from metallic LuNiSn-like matrix filled with semiconducting LuNiSb-like inclusions of the size of about 10 μm, appeared to exhibit metallic electrical resistivity and positive linear Seebeck coefficient. Nevertheless, this combination of the transport properties of LuNiSn and LuNiSb did not result in any increase of the thermoelectric power factor over that determined for the pure HH phase. Perhaps, a material with reversed microstructure pattern, i.e. HH semiconducting-like matrix and orthorhombic metallic inclusions with nanometric dimensions would possess better TE performance.

Regarding the LuNiSb compound, we found out that despite preparing its samples in the very same manner and using the same equipment, the physical properties show strong sample dependence. The observed diverse structural and transport behaviors of as-cast (not annealed) specimens probably arise from slightly different stoichiometry, structural disorder, and/or strain effects [30]. Possibly, a suitable annealing process would allow to obtain samples with similar physical characteristics.

**Acknowledgements**

This work was supported by the National Science Centre (Poland) under research grant no. 2015/18/A/ST3/00057.

**References**

[1] S. Chen, Z. Ren, Mater. Today 16 (2013) 387–395.
[2] J.-W.G. Bos, R.A. Downie, J. Phys. Condens. Matter 26 (2014) 433201.
[3] T. Zhu, C. Fu, H. Xie, Y. Liu, X. Zhao, Adv. Energy Mater. 5 (2015) 1500588.
[4] A. Page, P.F.P. Poudeu, C. Uher, J. Materiomics 2 (2016) 104–113.
[5] W.G. Zeier, J. Schmitt, G. Hautier, U. Aydemir, Z.M. Gibbs, C. Felser, G.J. Snyder, Nat. Rev. Mater. 1 (2016) 16032.
[6] S. Sportouch, P. Larson, M. Bastea, P. Brazist, J. Ireland, C.R. Kannewurf, S.D. Mahanti, C. Uher, M.G. Kanatzidis, MRS Proc. 545 (1998).
[7] K. Gofryk, D. Kaczorowski, T. Plackowski, A. Leithe-Jasper, Y. Grin, Phys. Rev. B 72 (2005).
[8] K. Gofryk, D. Kaczorowski, A.L. Jasper, Y. Grin, in:, IEEE, 2005, pp. 380–383.
[9] T. Sekimoto, K. Kurosaki, H. Muta, S. Yamanaka, MRS Proc. 886 (2005).
[10] K. Gofryk, D. Kaczorowski, T. Plackowski, J. Mucha, A. Leithe-Jasper, W. Schnelle, Y. Grin, Phys. Rev. B 75 (2007).
[11] T. Sekimoto, K. Kurosaki, H. Muta, S. Yamanaka, J. Appl. Phys. 102 (2007) 023705.
[12] K. Kawano, K. Kurosaki, T. Sekimoto, H. Muta, S. Yamanaka, Appl. Phys. Lett. 91 (2007) 062115.
[13] K. Kawano, K. Kurosaki, T. Sekimoto, H. Muta, S. Yamanaka, in:, IEEE, 2007, pp. 267–269.
[14] K. Kawano, K. Kurosaki, H. Muta, S. Yamanaka, J. Appl. Phys. 104 (2008) 013714.
[15] K. Gofryk, D. Kaczorowski, T. Plackowski, A. Leithe-Jasper, Y. Grin, Phys. Rev. B 84 (2011).
[16] A. Mukhopadhyay, S. Mahana, S. Chowki, D. Topwal, N. Mohapatra, in:, AIP Conference Proceedings, 2017, p. 110024.
[17] K. Synoradzki, K. Ciesielski, L. Kępiński, D. Kaczorowski, Phys. B Condens. Matter (2017).
[18] W. Xie, A. Weidenkaff, X. Tang, Q. Zhang, J. Poon, T. Tritt, Nanomaterials 2 (2012) 379–412.
[19] G. Tan, L.-D. Zhao, M.G. Kanatzidis, Chem. Rev. 116 (2016) 12123–12149.
[20] R.V. Skolozdra, A. Guzik, A.M. Goryn, J. Pierre, Acta Phys. Pol. A 92 (1997) 343–346.
[21] V.V. Romaka, L. Romaka, A. Horyn, P. Rogl, Y. Stadnyk, N. Melnychenko, M. Orlovskyy, V. Krayovskyy, J. Solid State Chem. 239 (2016) 145–152.
[22] J. Pierre, I. Karla, J. Magn. Magn. Mater. 217 (2000) 74–82.
[23] D.J. Bergman, L.G. Fel, J. Appl. Phys. 85 (1999) 8205–8216.
[24] J.E. Douglas, C.S. Birkel, M.-S. Miao, C.J. Torbet, G.D. Stucky, T.M. Pollock, R. Seshadri, Appl. Phys. Lett. 101 (2012) 183902.
[25] P. Sahoo, Y. Liu, J.P.A. Makongo, X.-L. Su, S.J. Kim, N. Takas, H. Chi, C. Uher, X. Pan, P.F.P. Poudeu, Nanoscale 5 (2013) 9419.
[26] A. Bhardwaj, N.S. Chauhan, B. Sancheti, G.N. Pandey, T.D. Senguttuvan, D.K. Misra, Phys. Chem. Chem. Phys. 17 (2015) 30090–30101.
[27] M.L.C. Buffon, G. Laurita, N. Verma, L. Lamontagne, L. Ghadbeigi, D.L. Lloyd, T.D. Sparks, T.M. Pollock, R. Seshadri, J. Appl. Phys. 120 (2016) 075104.
[28] A.E. Dwight, J. Common Met. 93 (1983) 411–413.
[29] V.V. Romaka, E.K. Hill, L. Romaka, D. Fruchart, A. Horyn, Chem Met Alloys 1 (2008) 298–302.
[30] H. Xie, H. Wang, C. Fu, Y. Liu, G.J. Snyder, X. Zhao, T. Zhu, Sci. Rep. 4 (2015).